# Temperature Dependence of Alkali-Metal Rattling Dynamics in the β-Pyrochlores, $AOs_2O_6$ (A = K, Rb, Cs), from MD Simulation

*Elvis Shoko[a), Vanessa K Peterson, and Gordon J Kearley*

*Australian Nuclear Science and Technology Organisation, Locked Bag 2001, Kirrawee DC, NSW 2232, Australia*

## Abstract

We investigate the temperature response of the alkali-metal rattling modes in β-pyrochlores, $AOs_2O_6$ (A = K, Rb, Cs), from the results of *ab initio* molecular dynamics (MD) simulations performed at 20 K, 100 K, and 300 K. Our results show that the temperature response of the $T_{1u}$ mode is clearly different from that of the $T_{2g}$ mode for all three pyrochlores. In this regard, two features are of particular note for both K and Rb; (1) the $T_{1u}$ mode exhibits a distinctly stronger softening response with decreasing temperature compared to the $T_{2g}$ mode, and (2) the $T_{1u}$ mode becomes stronger and sharper with decreasing temperature. These two findings suggest that the $T_{1u}$ mode is significantly more anharmonic and sensitive to the cage dynamics than the $T_{2g}$ mode. Examination of the local potentials around the alkali-metal atoms reveals that K has the flattest and most anharmonic potential at all temperatures while Cs exhibits the narrowest potential. The temperature dependence of the local potentials reveals that, for K, the potential at a higher temperature is not a simple extrapolation to higher energy of that at a lower temperature. Instead, we find significant reconstruction of the potential at different temperatures. Finally, we explore the temperature response of the coupling between the alkali metals and find a complex temperature dependence which suggests that the origin of the coupling may be more complex than a pure Coulomb interaction. We also find an unexpected increase in the static disorder of the system at low temperatures for the K and Rb pyrochlores.

# 1 Introduction

The β-pyrochlore osmates, $AOs_2O_6$ (A = K, Rb, Cs), are extensively studied because of their superconducting properties that are believed to be linked to the rattling modes of the alkali metals[1, 2]. The rattling of small atoms encaged in large cavities has been shown to be important for the thermoelectric performance of both clathrates[3] and skutterudites[4], and fuels our interest in the osmates. Although these compounds are metallic and therefore not suitable thermoelectric materials in their native form, by virtue of their simple composition, they provide a suitable system for investigating the fundamental physics of guest rattling in cage compounds. The β-pyrochlore osmates crystalize in cubic structures with space group $Fd\bar{3}m$ (No. 227)[5, 6], where *A*, Os, and O, occupy the 8b, 16c, and 48f sites, respectively. The alkali-metal atoms (rattlers) reside in a large cage formed by both O and Os atoms as illustrated in Figure 1. The alkali-metal atom can explore the large cage volume available to it which can lead to unusual

---

[a)] Author to whom correspondence should be addressed. Electronic mail: elvis.shoko@gmail.com. Telephone: +61-41-506-4823

dynamics. The dynamics get more complex as the ratio of the encaged alkali-metal atom volume to the cage volume decreases. For the K-Cs series, this ratio decreases in the order, Cs, Rb, and K, so that the K dynamics are expected to be the most complex in this series. Indeed, it has been found that the vibrational dynamics of the K atom in $KOs_2O_6$ differ significantly from those of Rb and Cs in their respective pyrochlores[7] and a complex low-energy signature was recently observed using inelastic neutron scattering (INS)[8].

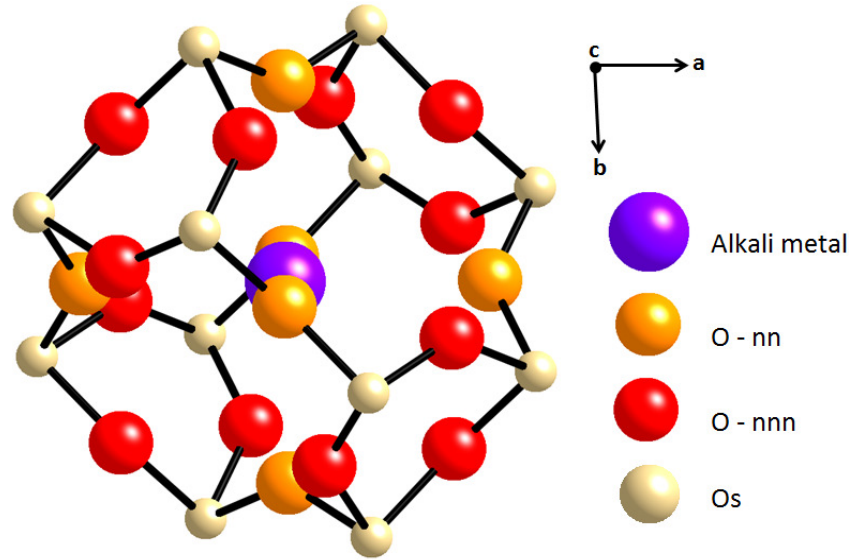

**Fig. 1** Illustration of the cage structure in the defect pyrochlore osmates, $AOs_2O_6$. The $Os_{12}O_{18}$ cage consists of six nearest neighbour O atoms (O – nn) octahedrally arranged around the cage centre followed by twelve next-nearest neighbour O atoms (O – nnn). Also forming the cage framework are twelve Os atoms as shown and the alkali metal atom occupies the cage center. The coordinate axes refer to the lattice vectors of the conventional unit cell (see Figure 2)

These dynamics are driven by highly anharmonic potentials presented by the cage environments around the encaged rattler atoms. Evidence for the anharmonicity of these potentials has been obtained from both theoretical and experimental approaches. We review the key results from the study of anharmonicity through the temperature dependence of some properties of the rattling dynamics. In theoretical models[9-12], anharmonicity has been described by expanding the potential energy up to fourth-order in the ion displacements, with[10, 11, 13] or without[9] the third-order term. Hattori and Tsunetsugu[10, 11] reported that the third-order term was particularly important for describing the anharmonic properties in $KOs_2O_6$ and they used their model to calculate the phonon softening with cooling which was reported from experiments[14, 15]. An inelastic neutron scattering (INS) experiment reported a softening of the alkali-metal modes with decreasing temperature[15] similar to the temperature-dependent phonon frequencies observed in the anharmonic filled skutterudites, e.g., $PrOs_4Sb_{12}$. In the $PrOs_4Sb_{12}$ example, the guest oscillation frequency decreased from 3.4 meV at 300 K to 2.4 meV at 10 K[16]. In addition, mean-square displacements from neutron diffraction have revealed temperature dependencies in all three β-pyrochlores[11] consistent with similar results for the anharmonic guests in filled skutterudites[17]. Elastic neutron scattering data show a concave temperature dependence of the oscillation amplitude of the K in $KOs_2O_6$ reflecting strong anharmonicity[18, 19]. Lastly, anomalous temperature dependence of the electrical resistivity, attributed to anharmonicity, was reported for the these β-pyrochlore osmates[20, 21].

In this study, we furnish further details of the microscopic picture of the anharmonicity by investigating the temperature dependence of the alkali-metal dynamics in all the three pyrochlores using *ab initio* molecular dynamics (MD) simulations. For each pyrochlore, we performed MD simulations at three different temperatures, i.e., 20 K, 100 K, and 300 K. The results at 300 K were validated against experimental INS data[22], whereas a similar validation could not be performed for the other temperatures as there are no published experimental INS data. We explore the rattler spectral properties along with the corresponding atomic root mean-square displacements (RMSDs) at the different temperatures and interpret these in terms of the local potentials around the rattlers. We find significant temperature dependence of the spectra, RMSDs, and local potentials confirming that the rattler dynamics involve highly anharmonic large-amplitude oscillations. Furthermore, $KOs_2O_6$ exhibits the most pronounced temperature dependence of the calculated local potentials around the K atoms. The alkali-metal sublattice coupling has been reported[23] to control the dominant alkali-metal dynamics at 300 K but the temperature dependence of this phenomenon has never been investigated. Consequently, we examine this issue and show that the alkali-metal sublattice coupling exhibits a complex temperature dependence partly influenced by a potential energy landscape of increasing multiplicity of local minima with decreasing temperature.

## 2    Computational Details

We performed low-precision *ab initio* MD simulations using the Vienna Ab initio Simulation Package (VASP)[24, 25] on the conventional cubic unit cell of the β-pyrochlore osmates $AOs_2O_6$ (A = K, Rb, Cs) consisting of 72 atoms, Figure 2. The Projector-Augmented Wave (PAW) method[26, 27] was employed and for the exchange-correlation potential, the generalized gradient approximation with the Perdew, Burke, and Ernzerhof (GGA-PBE) functional was used[28, 29]. The calculations were performed at the gamma point with a plane-wave cutoff energy of 300 eV.

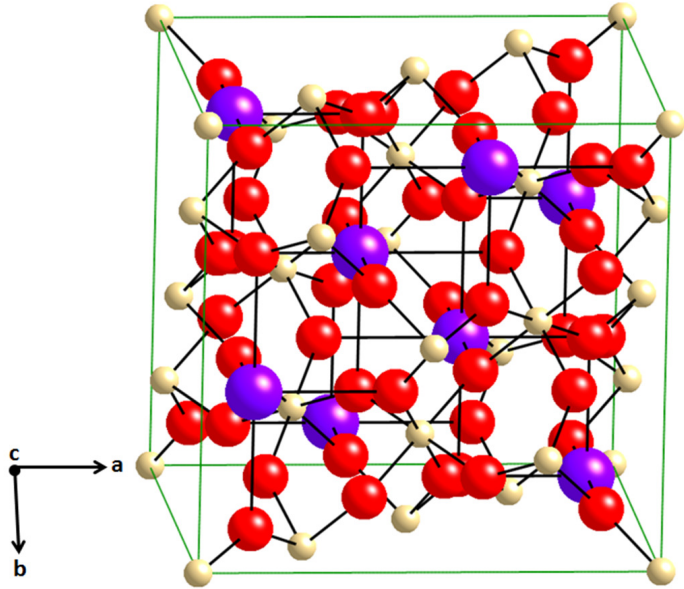

**Fig. 2** The conventional cubic unit cell of the defect-pyrochlore osmates used in the MD simulations. The cell consists of 72 atoms shown here as large violet (alkali metal atoms), medium-size red (O atoms), and small yellow (Os atoms) spheres.

For $KOs_2O_6$ we used the 300 K experimental lattice constant of 10.089 Å for all simulations at the different temperatures. The measured lattice constant[5, 30] for this compound shrinks slightly from 10.089 Å to 10.083 Å at 100 K and expands slightly to 10.093 Å at 20 K. For $RbOs_2O_6$, the experimental lattice constant of 10.1093 Å at 100 K was used for all simulations although the exact lattice constants are 10.1139 Å (300 K) and 10.1081 Å (20 K)[6]. From our previous work[31], we could not establish thermal stability for the microcanonical ensemble (NVE) MD simulation of the Al-doped $CsW_2O_6$ at the experimental lattice constant. For this reason, the $CsOs_2O_6$ lattice constant was calculated in VASP starting from the experimental value[32] of 10.1525 Å (Hiroi *et al.*[33] reported 10.1477 Å) by performing a relaxation permitting both cell shape and volume changes and the value 10.262 Å was obtained which we used for all the MD simulations. For all simulations, starting from the experimental atomic positions, the structures were first converged to < 0.001 eV/Å total forces per atom followed by equilibration (NVT) for 4 ps at the target temperatures (20 K, 100 K, and 300 K) and microcanonical ensemble (NVE) production runs of 36 ps were then performed. To facilitate direct comparison with experimental INS, we calculated simulated spectra in nMOLDYN[34] for the region in **Q**-ω phase space[35] (**Q** = neutron scattering vector, ω = frequency, representing the neutron energy through $E = \hbar\omega$) corresponding to the IN6 instrument data of Mutka *et al.*[15] used to validate our MD results at 300 K.

## 3  Results and Discussion

Due to the inherent rounding errors of the numerical algorithms, the simulations tended to either gradually cool down or heat up. In Table 1 we summarize the temperature statistics of the different simulations and the results show an overall decrease in variation down the K-Cs series. For all pyrochlores, the mean temperatures for the 20 K simulations are higher than target, whereas the opposite is true at 300 K. At 100 K, the Cs average is below target while both K and Rb are above. Overall, the mean temperatures show a decreasing trend from K to Cs at both 20 K and 100 K while the same trend does not hold at 300 K, with Cs higher than expected. The increase in the standard deviation with decreasing rattler size may be related to the greater flexibility of motion with smaller rattlers.

Table 1 MD simulation temperature statistics (K) for the defect pyrochlores. The asterisk (*) denotes missing data. The variation in temperature increases with decreasing rattler size which may imply greater flexibility of motion for smaller rattlers.

| | $KOs_2O_6$ Simulations | | |
|---|---|---|---|
| Target Temp | 20 | 100 | 300 |
| Mean | 30 | 112 | 263 |
| Std. Dev. | 6 | * | 27 |
| Range | 17-49 | * | 181-357 |
| | $RbOs_2O_6$ Simulations | | |
| Target Temp | 20 | 100 | 300 |
| Mean | 29 | 103 | 260 |
| Std. Dev. | 5 | 8 | 22 |
| Range | 17-45 | 76-135 | 185-336 |
| | $CsOs_2O_6$ Simulations | | |
| Target Temp | 20 | 100 | 300 |
| Mean | 24 | 80 | 276 |
| Std. Dev. | 3 | 11 | 18 |
| Range | 16-33 | 52-116 | 201-350 |

## 3.1 Elemental Spectra

As we are only interested in examining the rattling dynamics of the alkali metals and their dependence on temperature, the discussion of the spectral properties is restricted to the elemental spectra of these atoms only.

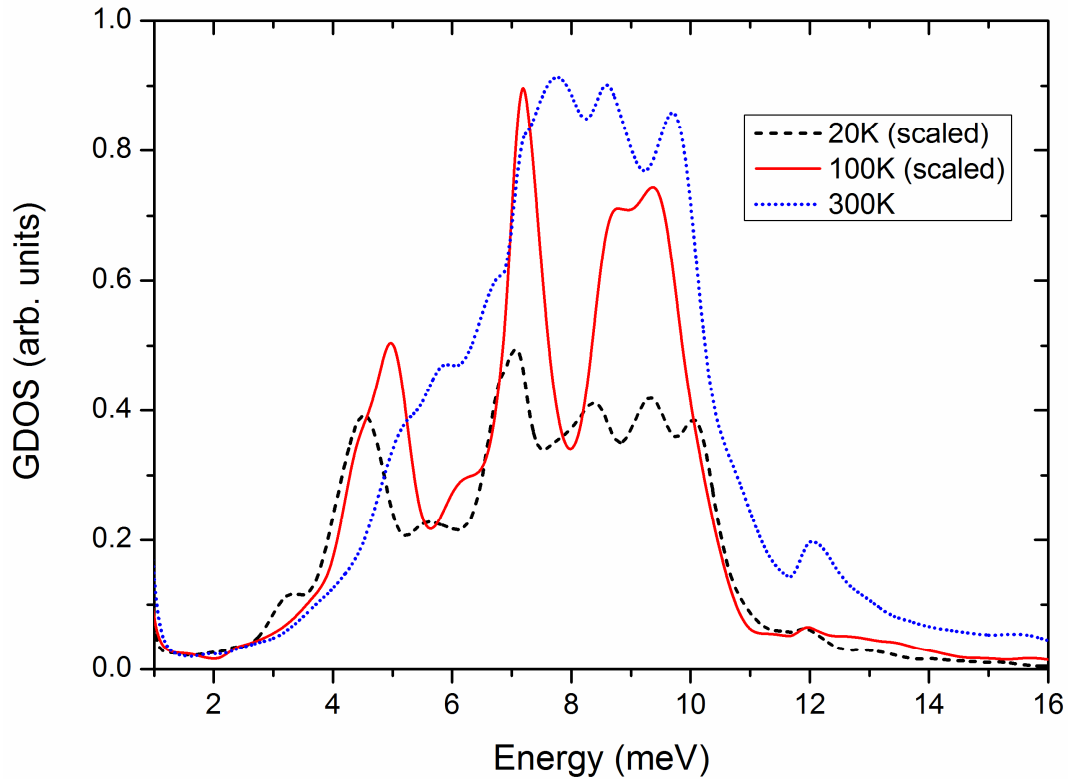

**Fig. 3** Frequency spectra of the K atoms at 20 K, 100 K, and 300 K showing that the vibrational modes are sensitive to temperature. To facilitate visual comparison, the magnitudes are scaled for 20 K (x4) and 100 K (x1.3). At 300 K, the spectrum exhibits a broad feature centered at about 8.6 meV consisting of three peaks with the upper peak at 9.7 meV. Shoulders appear below about 7.2 meV. At 100 K, the three-peak feature undergoes significant splitting into two prominent features; a peak at 7.2 meV and a feature centered at 9.1 meV consisting of two peaks which are poorly resolved. The upper peak is at 9.4 meV indicating a softening of the $T_{2g}$ mode by about 0.3 meV compared to 300 K. More striking is the appearance of the sharp $T_{1u}$ mode at 4.9 meV which was only a shoulder at 300 K. This peak softens by about 0.4 meV at 20 K and its intensity relative to the $T_{2g}$ mode has increased compared to 100 K. There is virtually no softening of the $T_{2g}$ from 100 K to 20 K.

To examine if there is evidence of softening of the alkali-metal modes with decreasing temperature we calculated the alkali-metal spectra from the MD simulations at 300 K, 100 K, and 20 K and these are plotted in Figures 3-5. In assigning the peaks in Figures 3-5 to the $T_{1u}$ and $T_{2g}$ modes, we follow previous work[7, 14, 36] together with our recent study[23]. Briefly, in this assignment of the alkali-metal modes, there are two main features of the spectrum; the low-energy feature assigned to the $T_{1u}$ mode, and a high-energy feature corresponding to the $T_{2g}$ mode. Figure 3 shows that significant changes occur to both the $T_{1u}$ and $T_{2g}$ K modes with changing temperature. The $T_{2g}$ mode consists of a broad feature made up of three poorly resolved peaks at 300 K with the upper peak estimated at 9.7 meV. At 100 K, significant splitting of the $T_{2g}$ mode occurs giving a prominent low-energy peak at 7.2 meV and a feature at higher energy centered at 9.1 meV. This feature consists of two poorly-resolved peaks, the upper one centered at 9.4 meV. Comparing this to the $T_{2g}$ upper peak at 300 K reveals a softening of about 0.3 meV from 300 K to 100 K. No further softening of the $T_{2g}$ mode occurs at 20 K; instead, there is a further splitting of the mode into four peaks. Since the $T_{2g}$ mode can only split into a maximum of three peaks, the four peaks we obtain at 20 K may suggest a slightly different symmetry. We attribute this to a lower symmetry due to trapping of the atoms in local minima at this lower temperature as will be discussed in Section 3.3. Hasegawa *et al*. reported a shift of 1.1 meV of the $T_{2g}$ mode to lower energy when the temperature was changed from 300 K to 4 K with a linear dependence in this temperature range[14]. Such a linear temperature dependence corresponds to shifts of 0.3 and 0.7 meV between the 20 K and 100 K, and 100 K and 300 K spectra, respectively. Thus our MD results exhibit about 0.3 meV less softening of the $T_{2g}$ mode compared to these data. In better agreement with our findings are the INS results[15], where virtually no softening of the $T_{2g}$ mode was reported; instead, the intensity of this mode decreased rapidly with decreasing temperature making it hard to locate it at lower temperatures ($\sim T < 150\ K$). Our results reveal the same general trend, with the intensity of the $T_{2g}$ mode relative to the $T_{1u}$ having decreased very significantly at 20 K compared to the other temperatures. Comparison could also be made to the results of a theoretical calculation[11] where softening of the K mode was reported but references to the $T_{1u}$ and $T_{2g}$ modes become meaningless for that work.

A most striking feature of Figure 3 relates to the $T_{1u}$ mode. In contrast to the $T_{2g}$ mode, the $T_{1u}$ mode shows much stronger softening with decreasing temperature. Although it is difficult to assign this mode at 300 K as it falls in the shoulder region of the spectrum, static cage calculations[23] show a sharp $T_{1u}$ mode at 6.6 meV. Compared to this value, Figure 3 shows that at 100 K, the $T_{1u}$ mode shifts by about 1.7 meV to 4.9 meV which is more than five times that of the $T_{2g}$ mode. A further softening of 0.4 meV occurs at 20 K, so that overall, the $T_{1u}$ more is more temperature-sensitive than the $T_{2g}$ mode. A similar steep temperature dependence of the $T_{1u}$ mode was reported from experimental INS data[15] and was reported to saturate at T ≤ 40 K. A theoretical calculation also reported strong softening of the $T_{1u}$ mode[37]. We pointed out elsewhere[23] that the $T_{1u}$ mode couples to the cage modes more strongly than does the K $T_{2g}$ mode and the present results appear to support that finding. Our results suggest that with the dynamics of the cage framework reduced at lower temperatures, the coupling to the $T_{1u}$ mode becomes weaker leading to a sharper $T_{1u}$ peak.

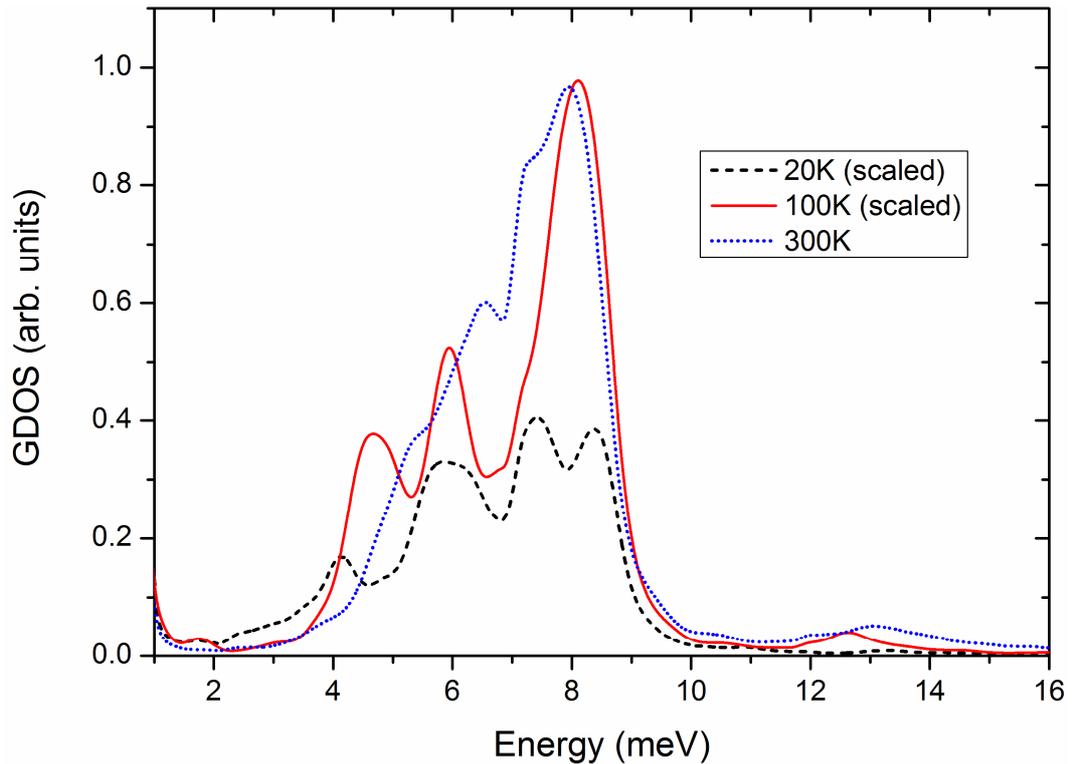

**Fig. 4** MD spectra of Rb in RbOs$_2$O$_6$ calculated at three different temperatures: 20 K, 100 K, and 300 K. To facilitate visual comparison, the magnitudes are scaled for 20 K (x6) and 100 K (x2.3). These spectra show an increasing splitting of the T$_{2g}$ mode with decreasing temperature as well as a T$_{1u}$ mode which is only a shoulder at 300 K becoming more well-defined a the lower temperatures. The T$_{1u}$ mode shows a strong temperature dependence shifting to lower energies from about 5.4 meV at 300 K to 4.6 meV at 100 K, and then to 4.1 meV at 20 K.

Compared to the K spectra (Figure 3), the Rb spectra in Figure 4 are narrower and, at 300 K, show a prominent feature corresponding to the T$_{2g}$ mode, with the T$_{1u}$ mode only appearing as a shoulder at 5.4 meV. This energy of the Rb T$_{1u}$ mode coincides with that obtained from a static cage calculation[23]. As with the K case, the T$_{2g}$ mode undergoes significant splitting with decreasing temperature while the T$_{1u}$ mode changes from a mere shoulder at 300 K to a well-defined softer peak at 100 K, shifted by about 0.8 meV which further softens by about 0.5 meV at 20 K. However, in contrast to the K case, the Rb T$_{2g}$ mode shows virtually no softening with decreasing temperature. Although these results are similar to the INS report[15] in exhibiting less softening for Rb compared to K, there is also a difference worth noting. In the INS case[15], the trend in softening is similar between the T$_{1u}$ and T$_{2g}$ modes whereas our results show quite distinct behavior with the T$_{1u}$ mode showing significantly stronger softening. Although a Raman experiment[38] did not report on the softening of the T$_{1u}$ mode in RbOs$_2$O$_6$, the trend they found for the T$_{2g}$ mode is similar to the INS report[15]. Thus our results for the behavior of the T$_{2g}$ mode for Rb match neither experimental result but the reason for this is not clear.

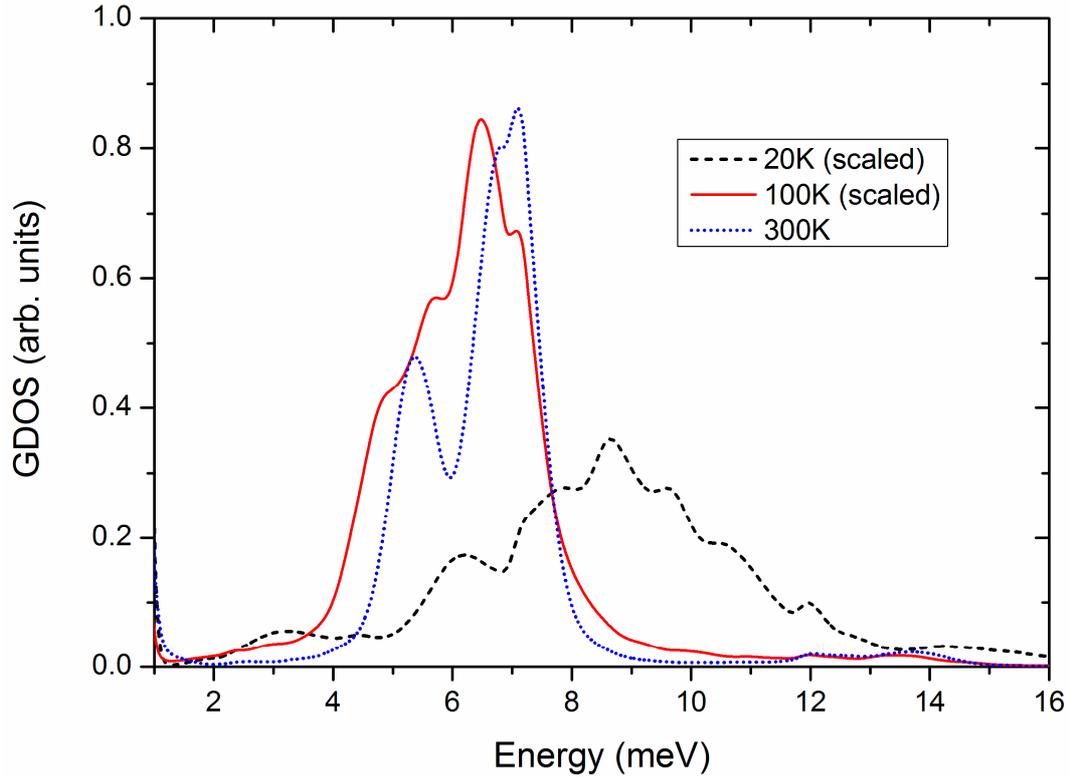

**Fig. 5** MD spectra of Cs in CsOs$_2$O$_6$ obtained at 20 K, 100 K, and 300 K. To facilitate visual comparison, the magnitudes are scaled for 20 K (x15) and 100 K (x3.5). The T$_{2g}$ peak at 7 meV at 300 K softens to 6.5 meV at 100 K, while the T$_{1u}$ mode at 5.3 meV for the 300 K spectrum becomes a shoulder. The 20 K spectrum exhibits an unexpected hardening with the main feature centered at 8.6 meV, while the low-energy feature appears at 6.2 meV.

The Cs spectra in Figure 5 exhibit several features which are quite distinct from either K or Rb. Whereas neither K nor Rb show a distinct T$_{1u}$ peak at 300 K, the Cs spectrum has a well-defined T$_{1u}$ peak at this temperature. This peak is less distinct at 20 K and nearly vanishes at 100 K, contrasting the K and Rb where the 100 K T$_{1u}$ peaks are the sharpest. Compared to 300 K, there is a small softening of the modes at 100 K, which is followed by an unexpected hardening at 20 K. Also unusual is the significant broadening of the T$_{2g}$ mode at this temperature. All these features seem to indicate that the low-temperature Cs dynamics approach the high-temperature K (and Rb) dynamics which is a puzzling result.

These MD results reveal two new important features of the rattler dynamics. Firstly, the T$_{1u}$ mode is almost non-existent at 300 K for both K and Rb but becomes well-defined at lower temperatures. In fact, for K, the evidence suggests a stronger T$_{1u}$ mode with decreasing temperature. This is consistent with a recent finding[23] that the T$_{1u}$ mode couples to the lattice more strongly than the T$_{2g}$ mode, such that it is damped out at the higher temperatures where the cage dynamics are stronger. Secondly, we find greater softening for the T$_{1u}$ mode with decreasing temperature compared to the T$_{2g}$ mode indicating that the two modes are somewhat distinct in nature, with the T$_{1u}$ more anharmonic. From a materials development and application perspective, the realization that the T$_{1u}$ and T$_{2g}$ modes exhibit different temperature responses may open an avenue for enhancing one mode over the other if that was desirable to achieve certain properties.

## 3.2 Root-mean-square Displacements (RMSDs) and Potential of Mean Force (PMF)

In this Section, we calculate the alkali-metal RMSDs for each pyrochlore at the three temperatures, 20 K, 100 K, and 300 K, and these values can be compared directly to the experimental data summarized by Hattori and Tsunetsugu[11]. We also calculate the potential of mean force (PMF) for the alkali metals at the different temperatures from the formula[39]:

$$PMF = -k_B T * \log P(\mathbf{r}) \quad (1)$$

Here, $k_B$ is the Boltzmann constant, $T$ is the absolute temperature, and $P(\mathbf{r})$ is the probability density function calculated from the trajectory population along the direction of maximum displacement, $\mathbf{r}$, in the principal axes coordinates. We fit the PMF to the harmonic potential model and departures from this model indicate anharmonicity of the local potential around an alkali metal atom which we quantify by calculating the reduced $\chi^2$ of the curve fitting from:

$$reduced\ \chi^2 = \frac{\chi^2}{DOF} = \frac{1}{DOF} \sum_i \left( \frac{p_i - h(r_i)}{\sigma_i} \right)^2 \quad (2)$$

Where $DOF$ = degrees of freedom, $p_i$ and $h(r_i)$ are the values of the PMF and harmonic potential at point $r_i$, and $\sigma_i$ is the standard deviation of $p_i$ which is set to unity since there are no error bars for $p_i$. Since the atoms traverse larger displacement ranges with increasing temperature, comparison of the reduced $\chi^2$ values for PMFs at different temperatures is not very meaningful as PMFs a higher temperatures would exhibit more scatter because of fitting over a larger range, hence larger $\chi^2$ values. To compare the PMFs at different temperatures for a given pyrochlore, we calculate the extrinsic curvature of each harmonic fit at the equilibrium (time-average) position which gives indication of the flatness of the bottom of the fitted model potential.

The mean square displacement for a harmonic oscillator is given by:

$$\langle u^2 \rangle = u_0^2 + \frac{\hbar}{2m\omega_E} \coth\left( \frac{\omega_E}{2k_B T} \right) \quad (3)$$

where $u_0$ is a temperature-independent parameter usually representing the static disorder of the system, $\hbar$ is the reduced Plank's constant, $m$, and $\omega_E$ are the mass, and characteristic energy of the oscillator, respectively. Equation (3) has been used to assess anharmonicity in the compounds of this study by comparing the values predicted from (3) to the experimental data[5, 18]. Since equation (3) assumes a single characteristic energy ($\omega_E$) and our results, Figures 3-5, show at least two characteristic energies, we will not use this relationship here. Instead, it suffices to compare our MD results to the published experimental values. We note that the RMSD values calculated from the MD do not contain $u_0$ and, as a result, they are generally expected to be lower than the experimental values, especially for systems with large static disorder.

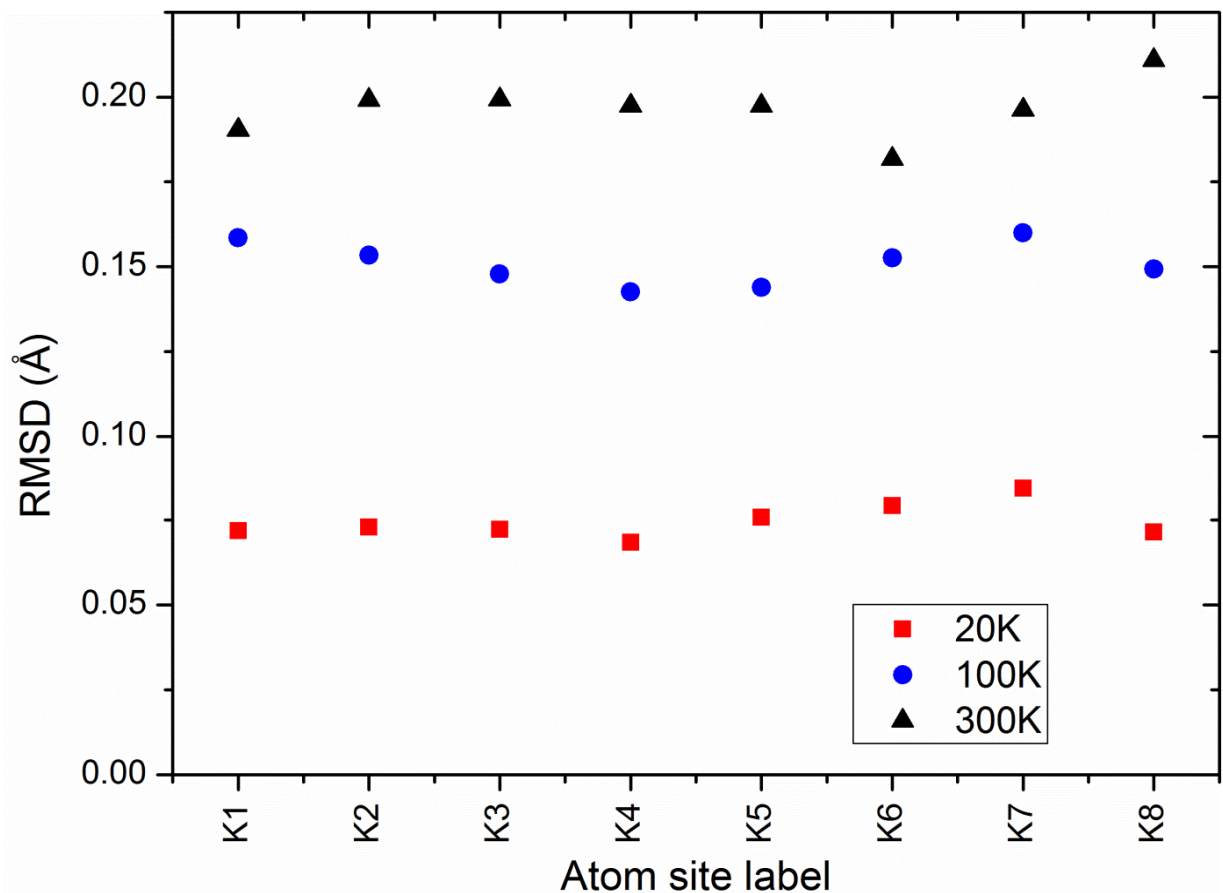

**Fig. 6** The effect of temperature on the RMSDs of the K atoms calculated from MD simulations at 20 K, 100 K, and 300 K. The atom site labels refer to the eight K atoms in the simulation cell. The RMSDs decrease with decreasing temperature and show reasonable comparison with experiment (see text). The site-to-site variation in the RMSDs does not show significant temperature dependence.

Figure 6 shows the temperature dependence of the RMSDs of the K atoms; the average values are 0.07, 0.15, and 0.20 Å for 20 K, 100 K, and 300 K, respectively which should be compared to the experimental values of 0.14, 0.18, and 0.25 Å. Although still lower than experiment, the MD values at 100 K and 300 K show better agreement than the value at 20 K, which is notably lower. We will show, in Section 3.3, that the discrepancy at 20 K is removed by taking into account the static disorder of the system.

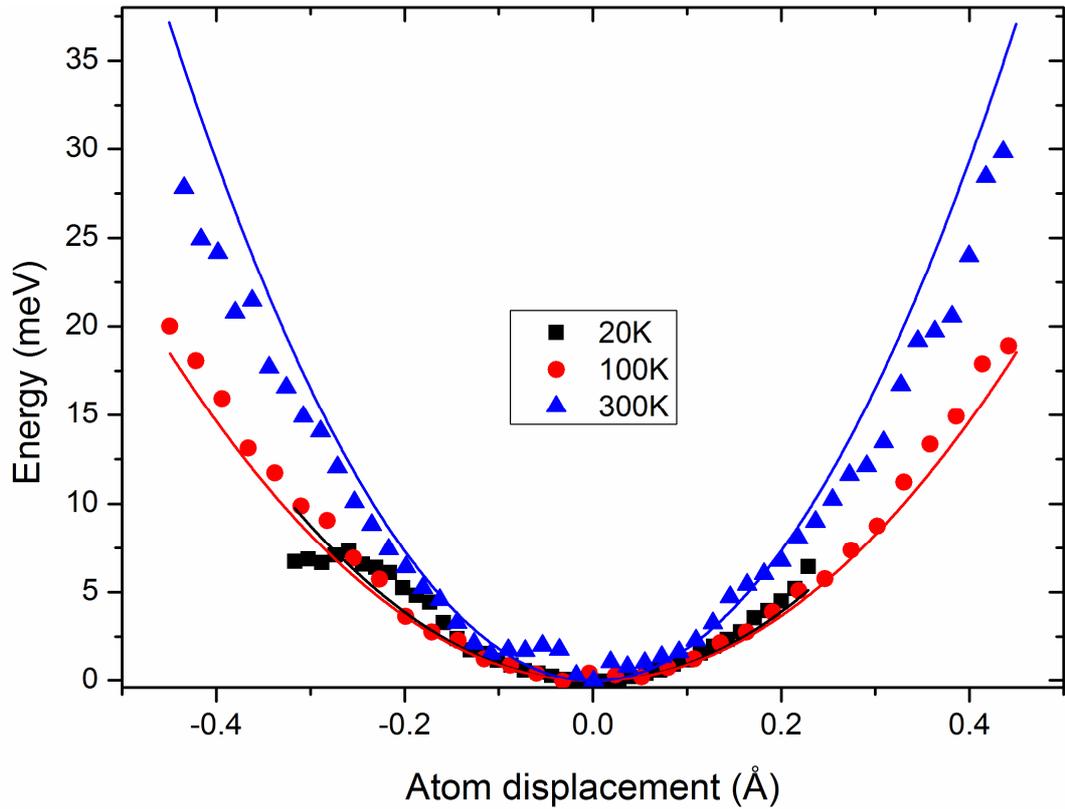

**Fig. 7** The average PMFs of the K atoms calculated from the MD simulations at 20 K, 100 K, and 300 K plotted along the direction of maximum displacement in the principal axes coordinates. In order to highlight the low-energy region of the PMF, we have restricted the plot to displacements up to ±0.45 Å although the full ranges are ±0.53 and ±0.72 Å at 100 K and 300 K, respectively. The continuous lines are harmonic fits to the PMF data and the reduced $\chi^2$ values for the fits (see text) are 0.8, 2.3, and 13.7 for 20 K, 100 K, and 300 K, respectively, indicating anharmonicity at all temperatures. That the 100 K PMF is broader than that at 300 K is evidence of phonon softening with decreasing temperature which further confirms the anharmonicity.

Figure 7 shows the K PMFs calculated from the MD simulations at 20 K, 100 K, and 300 K along with fits to the harmonic potential giving reduced $\chi^2$ values of 0.8, 2.3, and 13.7, respectively, reflecting anharmonicity at all three temperatures. Part of the stronger anharmonicity at higher temperature results from fitting the harmonic oscillator model over larger atomic displacements; for example, if the curve fitting is restricted to ±0.45 Å, then $\chi^2$ values at 100 K and 300 K are reduced to 0.68 and 0.76, respectively, values lower than that at 20 K. Nonetheless, the temperature dependence of these departures from the harmonic oscillator model confirms anharmonicity of the K dynamics. The harmonic fits in Figure 7 have extrinsic curvatures of 195, 183, and 367 Å$^{-1}$ at the equilibrium positions for 20 K, 100 K, and 300 K, respectively, representing a two-fold tightening of the potential from 100 K to 300 K. The anomalous tightening of the potential from 100 K to 20 K is consistent with the unshifted $T_{2g}$ mode in Figure 3 at this temperature. The reason for the anomaly at 20 K is not clear but it could be a result of the trapping of atoms in local minima at this low temperature as discussed in Section 3.3. Notwithstanding the anomaly at 20 K, the 1D PMFs in Figure 7 exhibit a broadening with decreasing temperature indicating a softening of the phonon frequencies as has been reported from INS[15] and Raman measurements[14].

In Figure 8 we plot the RMSD results for Rb at the three temperatures: 20 K, 100 K, and 300 K giving average values of 0.06, 0.11, and 0.17 Å, respectively. These values agree well with the corresponding experimental results[11] of 0.07, 0.12, and 0.17 Å, respectively. Just as in the case of K, the RMSD values decrease with decreasing temperature but the K temperature response, undergoing a decrease of 0.18 Å over this temperature range, is much stronger compared to only 0.11 Å for Rb.

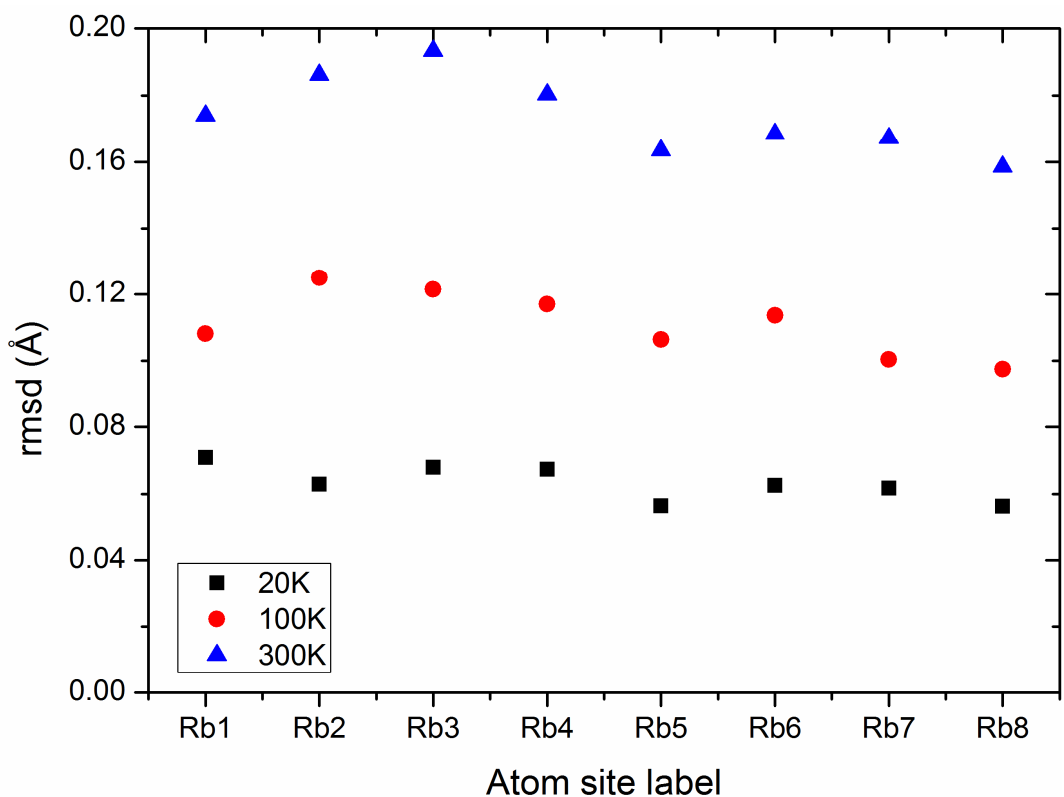

**Fig. 8** The Rb RMSD plotted at 20 K, 100 K, and 300 K showing a trend of decreasing RMSD values with temperature. The average values are 0.06, 0.11, and 0.17 Å at 20 K, 100 K, and 300 K, respectively which compare well with experiment (see text).

The 1D PMFs for Rb plotted in Figure 9 give fits to the harmonic oscillator model with $\chi^2$ values of 0.03, 1.09, and 3.86 at 20 K, 100 K, and 300 K, respectively. The extrinsic curvatures of the harmonic fits at the equilibrium positions are 276, 324, and 356 Å$^{-1}$ at 20 K, 100 K, and 300 K, respectively. These show a gradual flattening of the local potential around the Rb atoms with decreasing temperature. The extrinsic curvatures of both the Rb and K PMFs are comparable at 300 K but the K flattens out more rapidly at -0.92 Å$^{-1}$K$^{-1}$ to reach 183 Å$^{-1}$ at 100 K while the Rb is almost six times slower at -0.16 Å$^{-1}$K$^{-1}$. This demonstrates the strong anharmonicity in the K dynamics and indicates that the K and Rb dynamics are more different at 100 K than they are at 300 K.

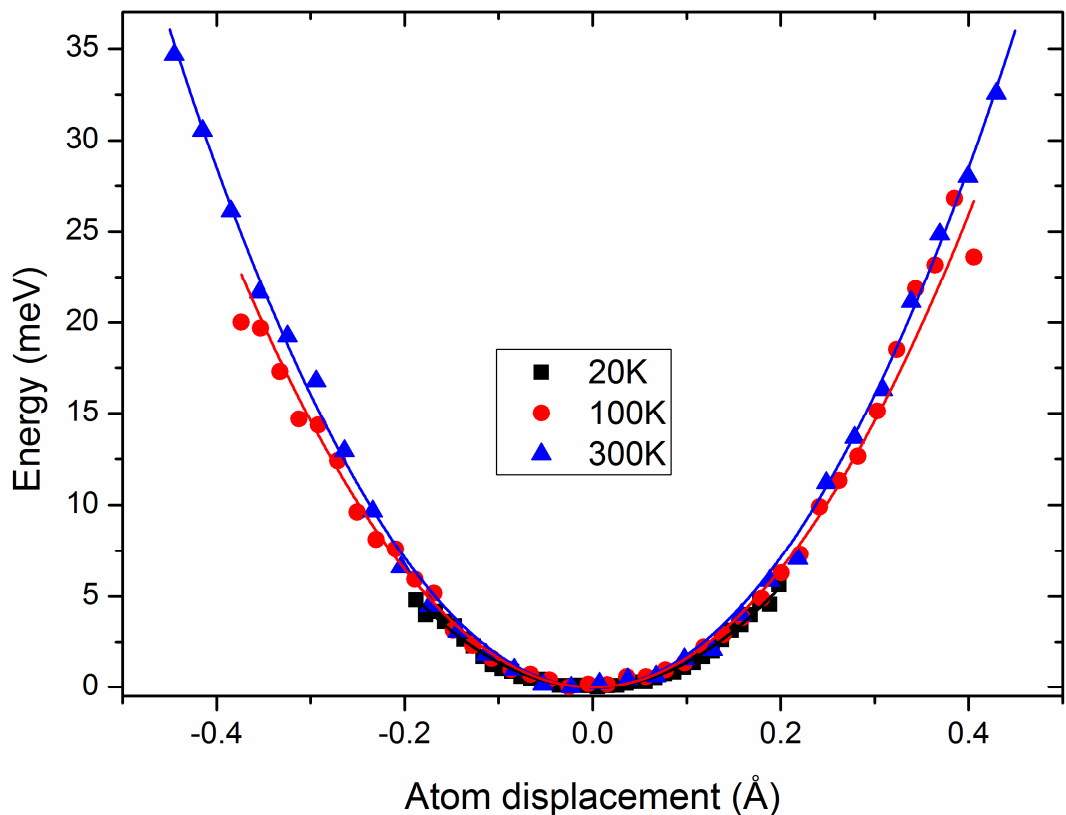

**Fig. 9** The Rb PMFs at different temperatures calculated along the direction of maximum displacement in the principal axes coordinates. Continuous lines represent fits to the harmonic oscillator model and for 300 K, with atom displacements up to ±0.58 Å, the plot is restricted to atom displacements within ±0.45 Å in order to highlight the low-energy region. The reduced $\chi^2$ values of the harmonic fits are 0.03, 1.09, and 3.86 at 20 K, 100 K, and 300 K, respectively.

Figure 10 shows the RMSDs plotted for each Cs atom in the simulation cell and the values decrease with decreasing temperature. The average values at 0.03, 0.10, and 0.13 Å are comparable to the corresponding experimental values[11] of 0.04, 0.07, and 0.13 Å at 20 K, 100 K, and 300 K, respectively.

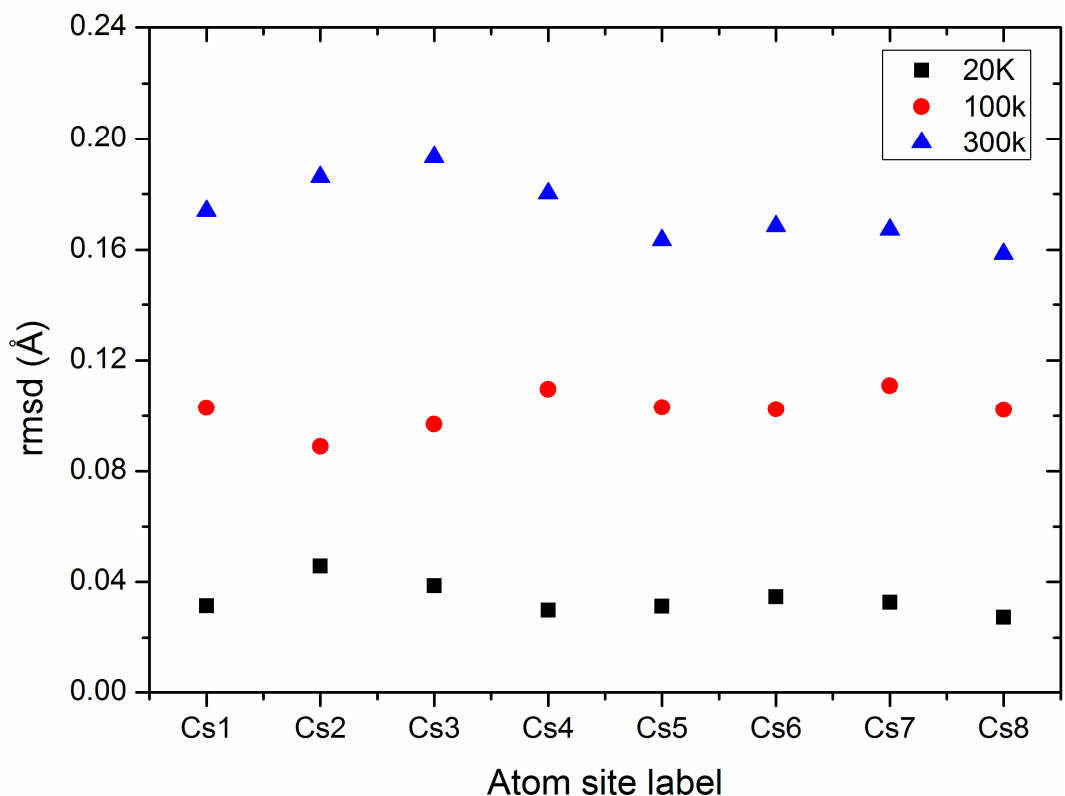

**Fig. 10** The RMSD of the Cs atoms calculated from the MD at different temperatures. The results show decreasing values with decreasing temperature. The MD average values of 0.03, 0.10, and 0.17 Å show reasonable agreement with experimental values[11] of 0.04, 0.07, and 0.13 Å at 20 K, 100 K, and 300 K, respectively.

In Figure 11, we plot the 1D PMFs for Cs and for each temperature we fit the data to the harmonic oscillator model obtaining reduced $\chi^2$ values of 0.47, 0.61, and 11.57 at 20 K, 100 K, and 300 K, respectively. It is surprising that the Cs anharmonicity is stronger than that of Rb at 20 K and 300 K, suggesting no systematic trend of decreasing anharmonicity with atomic number in the K-Cs series. It is also possible that the absence of such a systematic trend in these results reflects the shortcomings of performing the MD simulations for the different temperatures on a unit cell with the lattice constant for 300 K. The calculated extrinsic curvatures of the harmonic fits in Figure 11 are 553, 316, and 527 Å$^{-1}$, for 20 K, 100 K, and 300 K, respectively, giving the tightest local potential at 20 K. Compared to both the K and Rb PMFs at 300 K, Cs exhibits the tightest local potential giving relatively sharp spectral features in Figure 5. In contrast, the tight Cs local potential at 20 K does not lead to sharp spectral features in Figure 5; instead, a significant shifting of the modes to higher frequencies is obtained.

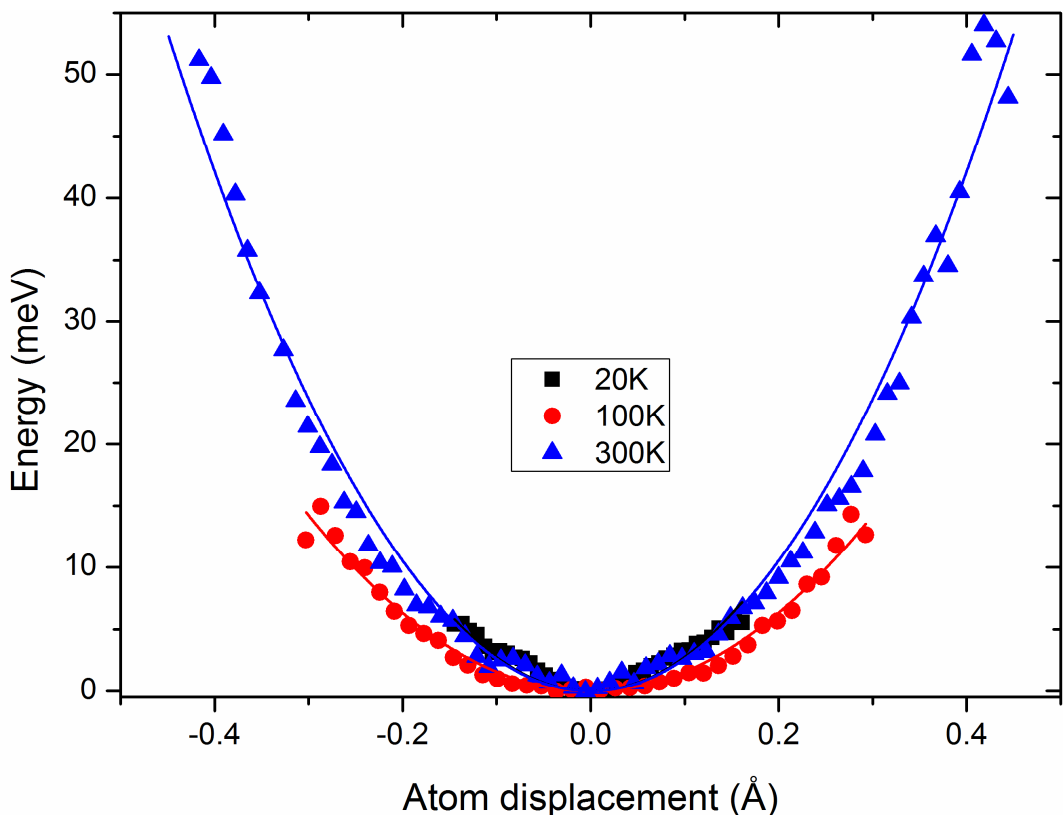

**Fig. 11** Cs PMFs at different temperatures with fits to the harmonic oscillator model (continuous lines). The reduced $\chi^2$ values of the harmonic oscillator fits are 0.5, 0.6, and 11.6 at 20 K, 100 K, and 300 k, respectively, a trend similar to both K and Rb.

Notwithstanding the shortcomings already noted from comparison to experiment, the study of the temperature dependence of the alkali-metal PMFs performed here reveals features crucial to understanding rattler dynamics in these compounds. In discussing the unusual temperature dependence of the anharmonicity in $KOs_2O_6$, Mutka *et al.* conjectured a potential with a flat bottom and steep walls which would permit phonon hardening with increasing displacement (temperature)[15]. A somewhat similar potential had been reported from a 0 K DFT calculation[40]. However, our finite-temperature calculations, Figure 7, reveal that the behavior at higher temperature differs from the simplistic picture of larger displacements on the same potential energy surface obtained at low temperature. Instead, we find that the potential essentially reconstructs at different temperatures so that the potentials at higher temperatures are not simply extrapolations of those at lower temperatures[41].

### 3.3 Alkali-metal Sublattice Coupling

We have discussed elsewhere[23] the role played by the strong metal-metal coupling on the alkali-metal sublattice in the rattling dynamics of these atoms. Here, we examine how this coupling responds to changes in temperature by comparing, for each compound, the Pearson correlation coefficients at 20 K, 100 K, and 300 K. All the correlation coefficients are calculated relative to the alkali-metal atoms. For each pair of atoms, we calculate the dynamical correlation matrix in the principal axes basis from which we select the absolute value of the matrix element with the largest magnitude. In Figures 12-14, we plot these correlation coefficients as a function of the distance of the different atoms from the alkali-metal sites. Although our method results in double counting, this does not matter as any double counts map onto the same point in Figures 12-14. We will refer to correlations between alkali metal atoms as A-A correlations, and those between an alkali metal and a cage atom as A-cage.

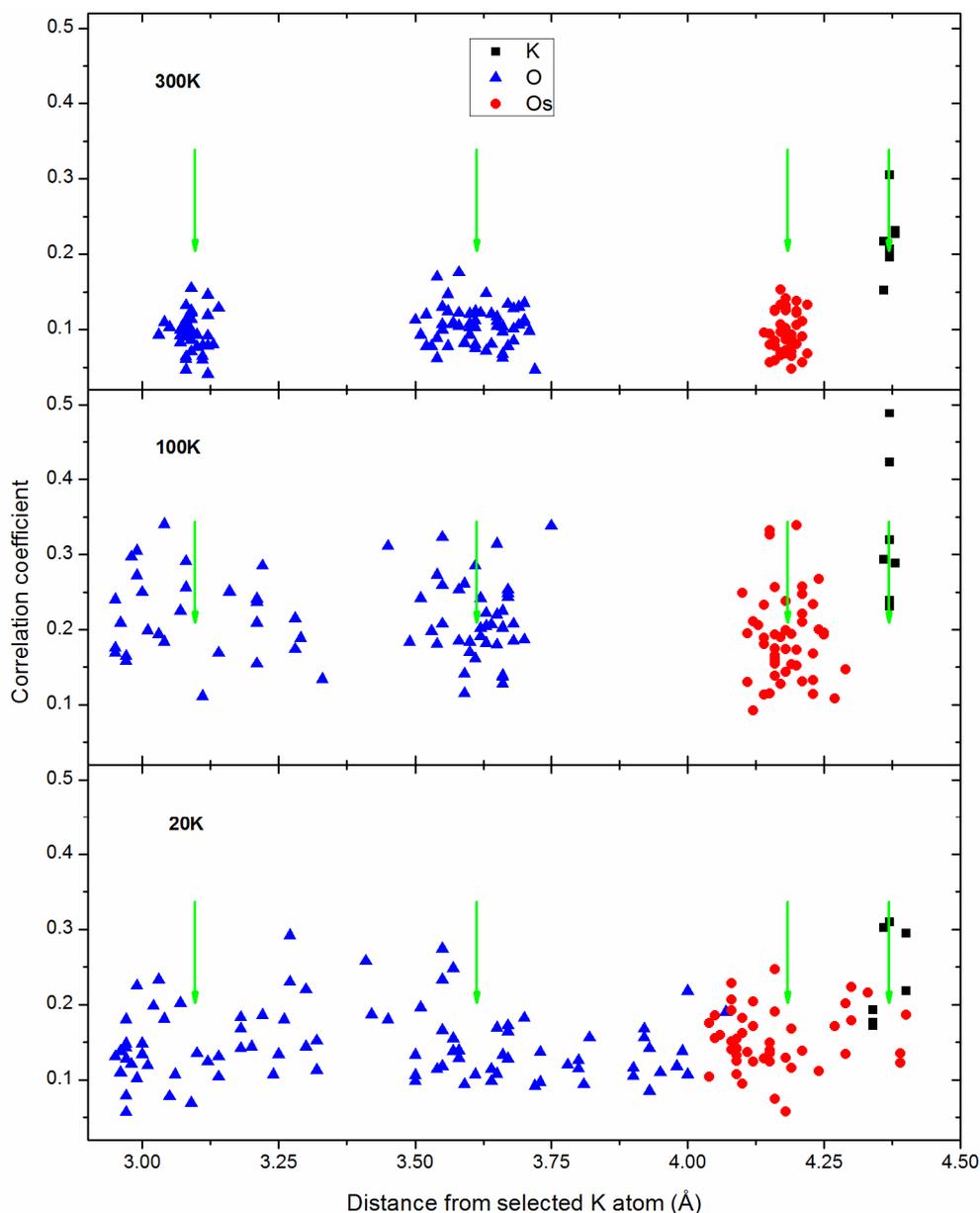

**Fig. 12** Pearson correlation coefficients with respect to K atoms in $KOs_2O_6$ calculated from MD simulations at 300 K, 100 K, and 20 K. The arrows indicate the crystallographic positions from experiment[5] with the two O clusters representing the nearest-neighbour (O-nn) and next-nearest-neighbour (O-nnn) coordination shells (see Figure 1). As the temperature decreases, the spread of the MD atomic positions around the experimental value increases probably a result of some trapping in local minima at these lower temperatures. Relative to the 300 K case, all the correlations at 100 K are higher. At 20 K, the correlations are lower than at 100 K but still higher than those at 300 K.

Even though the K-K distances are the largest, in all cases, the K-K correlations are higher than the K-cage correlations reflecting a strong K-K sublattice coupling.

Two general features can be noted from Figures 12-14. Firstly, the average A-A correlations are stronger than their A-cage counterparts for all the pyrochlores at all three temperatures included in this study, consistent with an earlier study of $KOs_2O_6$ at 0 K[42]. There appears to be no clear temperature-dependence trend for the A-A correlations; the correlations increase in the order 300 K, 20 K, and 100 K, for $KOs_2O_6$, 100 k, 20 K, and 300 K, for $RbOs_2O_6$, and for the $CsOs_2O_6$ case, they are virtually the same at 100 K as at 300 K but drop at 20 K. Secondly, the spread of the MD average atomic positions around the experimental crystallographic positions exhibit an increase with decreasing temperature, an effect that gets more pronounced with decreasing atomic number of the alkali-metal atom. This behavior probably reflects a proliferation of local minima with decreasing temperature in which the atoms are trapped because of the smaller RMSD values at low temperatures.

Experimentally, this effect could be observed by plotting the RMSD as a function of temperature and extrapolating to 0 K where a nonzero RMSD at this temperature indicates static disorder ($u_0$ in equation (3)). The most comprehensive data of this type has been published for $KOs_2O_6$ but, unfortunately, this compound undergoes an isomorphic phase transition at 9.5 K which seems to lead to a 0 K RMSD of 0.14 Å, slightly higher than the 0.12 Å obtained from extrapolation of the data before the phase transition. Nonetheless, the data give clear experimental evidence for static disorder in $KOs_2O_6$ which, as the MD shows, becomes important for the low-temperature atomic dynamics. Although the experimental data for Rb and Cs is not comprehensive enough to allow accurate extrapolation of the RMSDs to 0 K, the limited data indicates decreasing static disorder from $KOs_2O_6$ to $CsOs_2O_6$, with $RbOs_2O_6$ closer to $CsOs_2O_6$ than $KOs_2O_6$. Although the MD results show the same trend, the static disorder implied by Figures 12-14 suggest that the static disorder for $RbOs_2O_6$ is closer to that of $KOs_2O_6$ than $CsOs_2O_6$, and the reason for this is not clear.

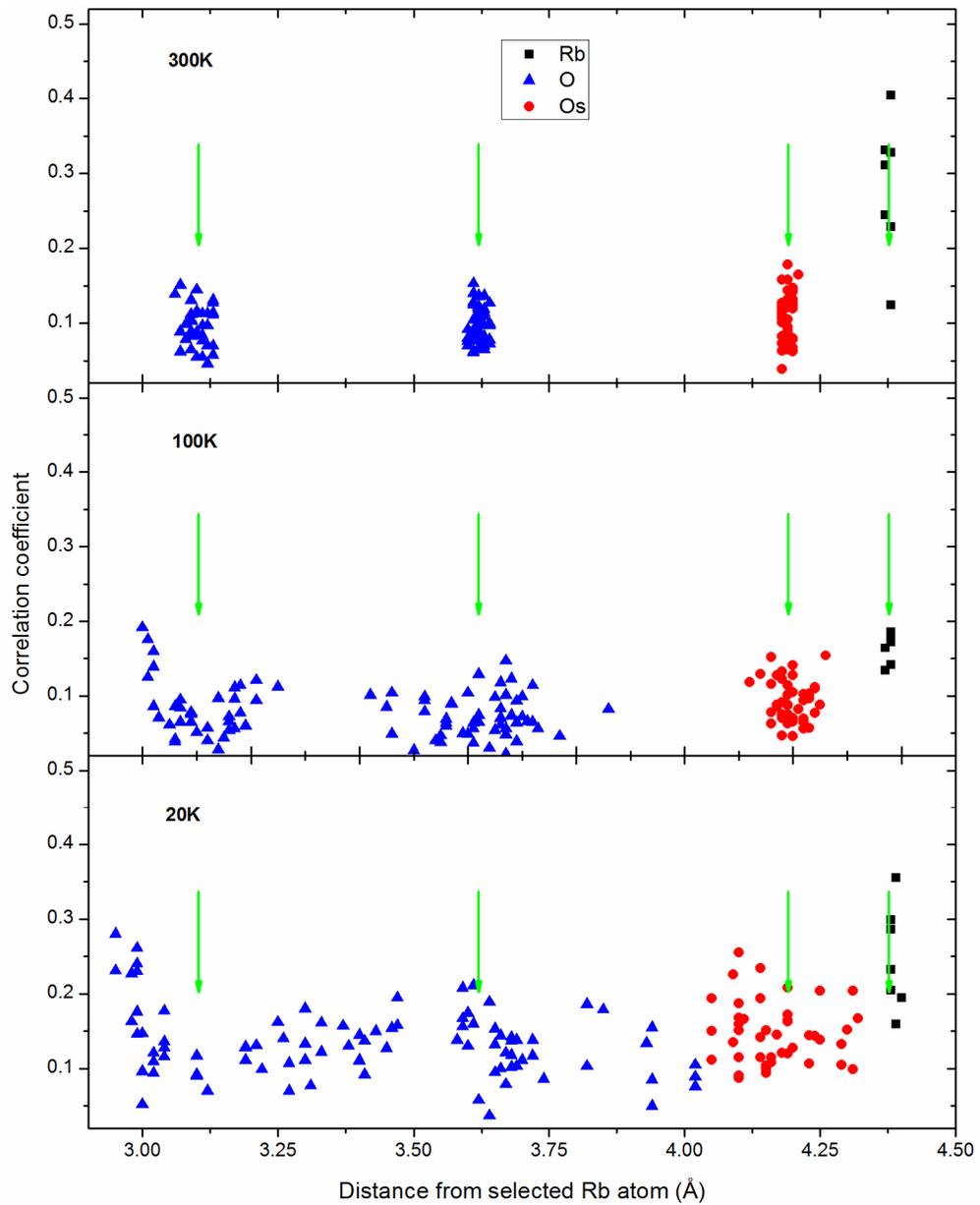

**Fig. 13** Pearson correlation coefficients relative to the Rb sites in RbOs$_2$O$_6$ calculated from MD simulations at 300 K, 100 K, and 20 K. The arrows indicate the crystallographic positions from experiment[6] with the two O clusters representing the nearest-neighbour (O-nn) and next-nearest-neighbour (O-nnn) coordination shells (see Figure 1). As in the KOs$_2$O$_6$ case, the spread of the MD atomic positions around the experimental value increases as the temperature decreases, probably a result of some trapping in local minima at these lower temperatures. At all temperatures, the Rb-Rb correlations are the highest indicating strong Rb-Rb coupling on the Rb sublattice. The Rb-Rb coupling strength decreases in the order 300 K, 20 K, and 100 K, respectively so that no clear trend can be established from these three temperatures.

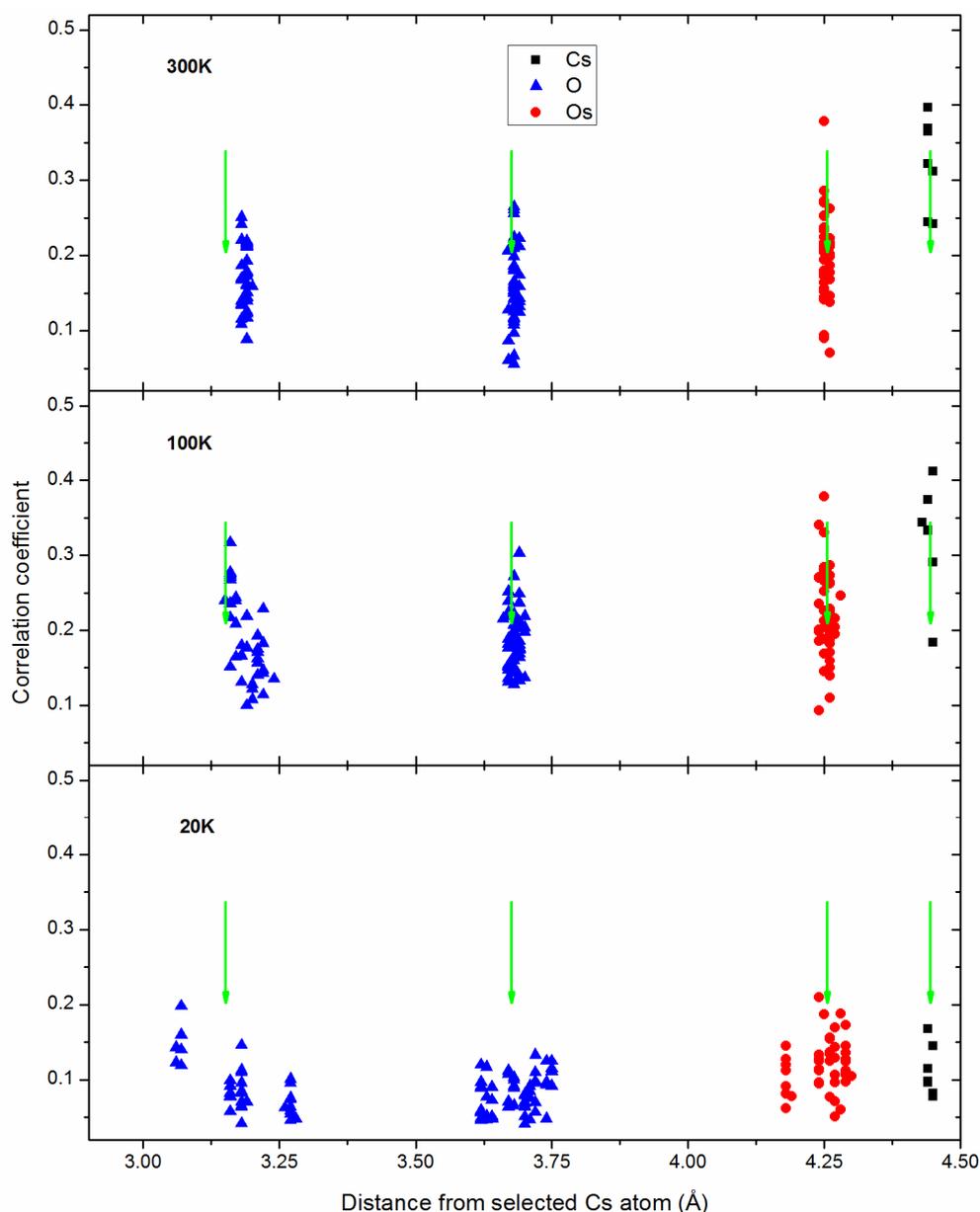

**Fig. 14** Pearson correlation coefficients of Cs in $CsOs_2O_6$ calculated from the MD simulation at 300 K, 100 K, and 20 K. The arrows indicate the crystallographic positions from experiment[43] with the two O clusters representing the nearest-neighbour (O-nn) and next-nearest-neighbour (O-nnn) coordination shells (see Figure 1). Although a spreading of the MD atomic positions around the experimental values occurs with decreasing temperature, the effect is far less pronounced than in other pyrochlores. This may indicate a potential-energy landscape with fewer local minima than in the other pyrochlores. The Cs-Cs correlations are higher than the Cs-cage correlations both at 300 k and 100 K, whereas at 20 K, the Cs-Os correlations become comparable. Even at 20 K, the Cs-Cs correlations should be considered higher when their farther distance is taking into account.

It is not clear whether the static disorder implied by Figures 12-14 is in the cage framework or the alkali-metal sublattice as the latter is the reference for calculating atom-atom distances. However, we can attempt to quantify and assign the static disorder by using as reference, the MD average structures at 300 K, since Figures 12-14 indicate negligible static disorder at this temperature for all the compounds. The procedure we adopt is as follows: At each temperature, time-average atomic positions are calculated for each compound. Then, for each element, a standard deviation is calculated from these positions relative to the corresponding positions at 300 K for all atoms of that element in the unit cell. This standard deviation is a measure of the static disorder for that element with reference to the 300 K average structure at that temperature. The results of this calculation are reported in columns 3-5 of Table 2, while in column 6 ('Lattice'), the combined average for the Os and O atoms is also included. Column 7 is the RMSD for the 'A' cation calculated from the MD as already discussed in Section 3.2, column 8 is the adjusted 'A' RMSD obtained by adding columns 3 and 7 together, i.e., a sum of the dynamic and static disorders for the 'A' cation, and column 9 is the RMSD for the 'A' cation from experiment[11] .

Table 2 shows that the static disorder does indeed increase with decreasing temperature and the increase is strongest for K and negligible for Cs. The results also reveal an interesting and unexpected feature of the static disorder: The alkali-metals exhibit the lowest static disorder compared to the cage framework, with the O atoms showing the highest disorder. That the cage framework gets more disordered with decreasing temperature than the rattlers is surprising because the cage framework is more rigid than the rattler sublattice. However, the simulation results suggest that the structure organizes itself around the rattler sublattice at low temperatures. This attests to the significant role played by the A-A coupling in controlling both the structure and dynamics in these defect pyrochlores, particularly at low temperatures. Column 8 shows that adding the static disorder to the 'A' cation RMSDs generally improves the agreement with experiment, particularly for the K at 20 K. We note that Yamaura *et al.* have suggested that the 0 K RMSD value of 0.14 Å for the K atoms should be attributed to the zero-point motion rather than static disorder[5] but our results suggest that there may be a significant static disorder component, as much as 50 % at 20 K, from these results. Further work is needed to ascertain that the static disorder we find at low temperatures is not simply an artefact arising from the trapping of atoms in local potentials at these low temperatures. This could be achieved by running much longer MD simulations or applying metadynamics to better sample the energy landscape. Alternatively, MD simulations where the atomic positions of the alkali-metal atoms are fixed at their values at 300 K could be performed to determine if the cage disorder still occurs for the low-temperature simulations.

Table 2. Using the MD average structures to quantify the static disorder with decreasing temperature in the β-pyrochlores. For each pyrochlore at each temperature, standard deviations are calculated relative to the MD average structure at 300 K. 'A' = K, Rb or Cs, and 'Lattice' refers to the cage framework. Column 7 is the calculated RMSD for the 'A' cation, column 8 is the sum of columns 3 and 7, while column 9 is the corresponding experimental data. For each pyrochlore, the 'A' cation exhibits the least static disorder, O the most, with Os intermediate. There is an approximate trend of decreasing disorder from K to Cs consistent with Figures 12-14.

| Pyrochlore | Temp (K) | Standard Deviation (Å) | | | | 'A' RMSD (Å) | Adj. 'A' RMSD (Å) | Exp. 'A' RMSD[11] (Å) |
|---|---|---|---|---|---|---|---|---|
| | | 'A' | Os | O | Lattice | | | |
| KOs$_2$O$_6$ | 100 | 0.0082 | 0.0279 | 0.0685 | 0.0609 | 0.15 | 0.16 | 0.18 |
| | 20 | 0.0747 | 0.0757 | 0.1646 | 0.1482 | 0.07 | 0.14 | 0.14 |
| RbOs$_2$O$_6$ | 100 | 0.0055 | 0.0321 | 0.0722 | 0.0644 | 0.11 | 0.12 | 0.12 |
| | 20 | 0.0111 | 0.0663 | 0.1496 | 0.1339 | 0.06 | 0.07 | 0.07 |
| CsOs$_2$O$_6$ | 100 | 0.0053 | 0.0065 | 0.0155 | 0.0138 | 0.10 | 0.11 | 0.07 |
| | 20 | 0.0053 | 0.0066 | 0.0124 | 0.0112 | 0.03 | 0.04 | 0.04 |

In Figure 15, we plot the A-A average correlations at each temperature as a function of distance up to 8.6 Å. Since periodic boundary conditions (PBCs) are applied in these simulations, correlations for separation distances above 5 Å may contain effects of periodic images and at least a 2x2x2 simulation supercell is required to properly treat them. However, since the compounds studied here have the same crystal structure, these effects are expected to be similar for all compounds. Thus we have included in Figure 15 correlations for distances above 5 Å only for comparisons across the K-Cs series where the effects of the PBCs can be neglected.

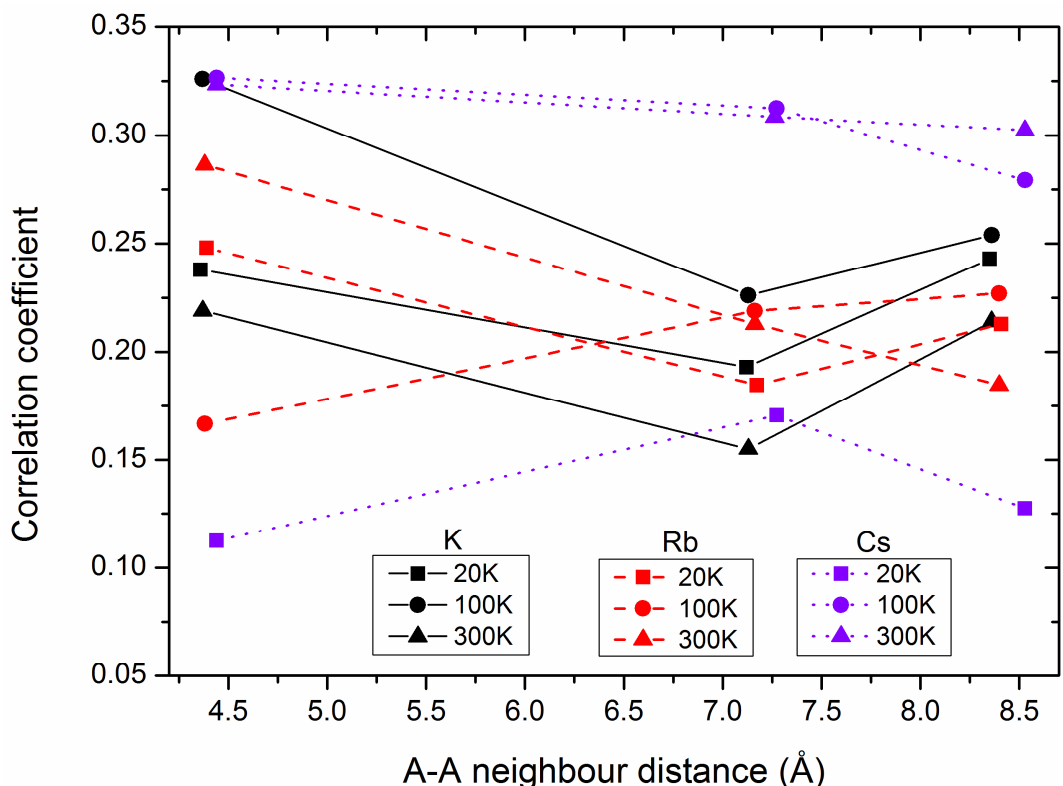

**Fig. 15** Alkali-metal sublattice coupling as measured by the dynamical correlation coefficients calculated from the MD simulations at 20 K, 100 K, and 300 K for all three pyrochlores. A-A in the abscissa label refers K-K, Rb-Rb, or Cs-Cs and the lines are a guide to the eye being continuous (black) for K, dashed (red) for Rb, and dotted (violet) for Cs. The different pyrochlores show a complex temperature-dependence for the A-A coupling. The A-A sublattice consists of 4 nearest neighbours, and 12 each for the next-nearest, and next-next-nearest neighbours. To properly extract correlations to the third nearest neighbours, a 2x2x2 supercell is required.

Figure 15 reveals different temperature response characteristics for the A-A coupling strength between the different pyrochlores. The Cs-Cs correlations at 100 K and 300 K are the highest showing only a slight decrease to the third nearest neighbours. The K-K correlations at 100 K are the next largest, but these together with those at 20 K and 300 K all drop at the second nearest-neighbour shell and then rise at the third shell. The same behavior is observed for Rb at 20 K suggesting that the broad characteristics of the Rb dynamics at this temperature resemble those of K as can also be seen by comparing the Rb 20 K spectrum (Figure 4) to those of K (Figure 3). This appears to be more complex behavior not readily understood in terms of the Coulomb interaction[42, 44]. At 300 K, the Rb correlations as a function of distance resemble Cs at the same temperature but fall off more sharply. Overall, these results show that, for a given pyrochlore, the A-A coupling undergoes complex changes with temperature so that there is no clear trend for the temperature dependence. Neither is it possible to make general statements about the strength of the A-A coupling across the series as has been assumed in previous work[42] since this depends on the particular temperature at which the comparison is made. Obviously our results obtained at only three temperatures and for the conventional unit cell are not sufficient to provide complete information for the temperature dependence of the A-A coupling. More data needs to be obtained at more temperatures and preferably for the larger 2x2x2 supercell to provide more robust conclusions about these trends and clarify whether the A-A interaction is purely Coulombic or more complex in nature.

# 4 Conclusion

For K and Rb, the alkali-metal modes soften with decreasing temperature with the effect being more pronounced for K, reflecting greater anharmonicity for $KOs_2O_6$. In both K and Rb, the shift of the $T_{2g}$ mode to lower energy with decreasing temperature is much less pronounced compared to experiment. These MD results uncover four significant features of the rattling dynamics. Firstly, the results show a distinct difference between the temperature response of the $T_{1u}$ mode and that of the $T_{2g}$. The $T_{1u}$ is strikingly more sensitive to the temperature, getting stronger and softer with decreasing temperature. The Cs modes exhibited a different temperature behavior compared to the other pyrochlores. Its $T_{1u}$ mode is sharpest at 300 K, almost vanishes at 100 K, and then emerges at 20 K but still much less pronounced than at 300 K. The Cs modes soften slightly from 300 K to 100 K before an unexpected hardening at 20 K. Secondly, the K PMFs at higher temperatures are not simply extrapolations to higher energy of the low-temperature PMF. Instead, the MD results show evidence of significant reconstruction at each temperature. The alkali-metal PMFs at the three temperatures exhibit the strongest anharmonicity and flatness of the potential for K while Cs shoes the narrowest potentials. Thirdly, the average MD structures show increasing static disorder with decreasing temperature, an effect that is particularly strong for $KOs_2O_6$, and suggests that at low temperatures, the crystal structure is organized around the alkali-metal sublattice. Further work is needed to ascertain that this effect is not simply a consequence of trapping of the atoms in local minima at low temperatures which could be overcome by running much longer MD simulations or other approaches[31]. Fourthly, we find that the A-A coupling exhibits more complex behavior with temperature suggesting the possibility that the A-A interaction may be more complex than a purely Coulombic interaction.

# 5 Acknowledgements

This work was partly supported by the Multi-modal Australian ScienceS Imaging and Visualisation Environment (MASSIVE) (www.massive.org.au).